\documentclass[a4paper,fleqn,usenatbib,useAMS]{mnras}

\usepackage[dvipsnames]{xcolor}
\usepackage{tikz,hyperref}

\definecolor{lime}{HTML}{A6CE39}
\DeclareRobustCommand{\orcidicon}{
	\begin{tikzpicture}
	\draw[lime, fill=lime] (0,0) 
	circle [radius=0.13] 
	node[white] {{\fontfamily{qag}\selectfont \tiny ID}};
	\draw[white, fill=white] (-0.0625,0.095) 
	circle [radius=0.007];
	\end{tikzpicture}
	\hspace{-2mm}
}

\foreach \x in {A, ..., Z}{\expandafter\xdef\csname orcid\x\endcsname{\noexpand\href{https://orcid.org/\csname orcidauthor\x\endcsname}
			{\noexpand\orcidicon}}
}

\usepackage{newtxtext,newtxmath}

\usepackage{booktabs}

\usepackage[T1]{fontenc}
\usepackage{ae,aecompl}

\usepackage{graphicx}	
\usepackage{amsmath}	
\usepackage{booktabs}
\usepackage{times}
\input{epsf}

\title[Morpho-kinematic model of R\,Aqr]
{Shaping the nebula around the symbiotic system R\,Aquarii} 
\author[E.~Santamar\'{i}a et al.]{E.~Santamar\'{i}a\thanks{E-mail:\,e.santamaria@irya.unam.mx}$^{1}\orcidA$, J.~A.~Toal\'{a}\thanks{Visiting astronomer at the Instituto de Astrof\'{i}sica de Andaluc\'{i}a (IAA-CSIC, Spain) as part of the Centro de Excelencia Severo Ochoa Visiting-Incoming
programme.}$^{1}\orcidC$, T.~Liimets$^{2}\orcidF$, M.~A.~Guerrero$^{3}\orcidB$, M.~K.~Botello$^{4}\orcidG$, L.~Sabin$^{4}\orcidE$
\newauthor and G.~Ramos-Larios$^{5}\orcidC$ 
\\
$^{1}$Instituto de Radioastronom\'{i}a y Astrof\'{i}sica, Universidad Nacional Aut\'{o}noma de M\'{e}xico, Morelia 58089, Michoac\'{a}n, Mexico\\
$^{2}$Astronomical Institute, Czech Academy of Sciences, Fri\v{c}ova 298, 25165 Ond\v{r}ejov, Czech Republic\\
$^{3}$Instituto de Astrof\'{\i}sica de Andaluc\'{\i}a, IAA-CSIC, Glorieta de la Astronom\'{\i}a s/n, 18008 Granada, Spain\\
$^{4}$Instituto de Astronom\'{i}a, Universidad Nacional Aut\'{o}noma de M\'{e}xico, Ensenada 22860, Baja California, Mexico\\
$^{5}$Instituto de Astronom\'{i}a y Meteorolog\'{i}a, CUCEI, Universidad de Guadalajara, Av. Vallarta 2602, Col. Arcos Vallarta, 44130 Guadalajara, Mexico 
}

\date{\today}

\pubyear{2024}

\begin{document}
\label{firstpage}
\pagerange{\pageref{firstpage}--\pageref{lastpage}}
\maketitle

\begin{abstract}
We present an analysis of high-dispersion spectroscopic observations of the symbiotic system R~Aquarii (R~Aqr) obtained with the Manchester Echelle Spectrograph (MES) at the 2.1 m telescope of the San Pedro M\'{a}rtir Observatory (Mexico) in conjunction with available narrow-band images. 
The data are interpreted by means of the {\sc shape} software to disclose the morpho-kinematics of the nebulosities associated with R~Aqr. 
The model that best reproduces narrow-band images and position-velocity diagrams consists of three structures: an outer (large) hourglass structure surrounding an inner bipolar with a spiral-like filament entwined around the later. 
The expansion velocity pattern of each structure is defined by different homologous expansion laws, which correspond to kinematic ages of $\tau_1$=450$\pm$25~yr (outer hourglass), $\tau_2$=270$\pm$20~yr (inner bipolar) and $\tau_3$=285$\pm$20~yr (spiral-like filament). 
We suggest that the spiral-like filament is tracing the regions of interaction of a precessing jet with the circumstellar material, which simultaneously carves the inner bipolar structure. 
If a similar process created the large hourglass structure, it means that the action of the jet ceased for about 170~yr. We discuss the implication for other unresolved symbiotic systems detected in X-rays.
\end{abstract}




\section{Introduction}
\label{sec:introduction}

The symbiotic system R~Aquarii (hereinafter R~Aqr) is one of the most studied objects of this class, from radio to the X-ray regime, including optical and UV studies \citep[e.g.,][]{Bujarrabal2021,Hollis1985,Hollis1997,Kellogg2001,Michalitsianos1980}. 
This symbiotic system includes an accreting white dwarf (WD) and a Mira-type star with a variability period of 387 d \citep{Gromadzki2009}. 
The orbital period of the system is 43.6~yr.

At a distance of $\sim$180 pc (\citealt{Solf1985}, \citealt{Liimets2018}), it is the closest known symbiotic system, which offers a unique opportunity to study its complex outflows  with unprecedented detail. One of the most notorious morphological feature in R~Aqr is the extended hourglass structure first reported by \citet{Lampland1922}. \citet{Liimets2018} offered an unparalleled  portrayal of the nebula using multi-filter Very Large Telescope (VLT) images obtained with the Focal Reducer/low-dispersion Spectrograph 2 (FORS2) that showed in great detail the multi layers of the hourglass structures best seen in the H$\alpha$+[N\,{\sc ii}] filter. Another main nebular feature is an S-shaped bipolar structure protruding from the central source, very likely the result of a precessing jet. This has also been the focus of numerous studies, but it was best imaged by the {\it Hubble Space Telescope (HST)} instruments \citep{Paresce1991,Paresce1994,Melnikov2018,Huang2023}.

The external hourglass structure has been found to expand ballistically and estimated to have an age of 685 yr when converting the literature ages (\citealt{Solf1985}, \citealt{Liimets2018}) into the present time. 
It has a knotty morphology and is more prominent in lower ionisation lines of H$\alpha+$[N\,{\sc ii}] and [O\,{\sc ii}], while in higher ionisation lines, such as [O\,{\sc iii}], it appears diffuse and fainter \citep[see Fig.~\ref{fig:slits};][]{Liimets2018}. The hourglass feature extends 2.5~arcmin in the east-west (EW) direction and $\approx$1.7~arcmin in the north-south (NS) direction. By modelling their long-slit spectral data, \citet{Solf1985} found that this bipolar structure is tilted by 18$^\circ$ with respect to the plane of the sky and that its equatorial waist is expanding at a velocity of 55 km s$^{-1}$. Overall, radial velocities (with respect to the systemic velocity of $-23$ km s$^{-1}$) are reported to range from  $-85$ to 75 km s$^{-1}$, increasing with increasing latitude angle of the expansion direction.

Additionally, the long-slit spectra of \cite{Solf1985} revealed the presence of the inner bipolar nebula, which shares the same polar-axis with the outer nebula as well as similar radial velocity values. In their model, the inner structure is smaller than the outer bipolar nebula. It is thinner in the EW direction ($\sim$25~arcsec) than in the NS direction (75~arcsec). Hence, it appears stretched in the NS direction compared to the outer hourglass structure which is flattened in the NS direction. They also find that the inner nebula shares the same inclination angle with respect to the plane of the sky as the outer nebula and that its equatorial waist expands at a velocity of 32 km~s$^{-1}$. Further work by \citet{1992A&A...257..228S} confirmed the presence of the two bipolar structures.

The jet-like feature, designated as a {\it spike} by \citet{Solf1985}, was considered to be formed by condensations of higher electron density areas in a thin surface layer of both nebulae rather than a distinct jet-like feature. \citet{1992A&A...257..228S} reported noticeable changes in the jet feature between data from 1982 and up to 7 yr later.
They propose that the jet-like condensations are formed due to the shocks by the highly supersonic WD wind impinging against the more slowly expanding inner bipolar shell. Simultaneously, they acknowledge the complexity of explaining significant differences in line widths among various jet-like features.

A second modelling of the complex outflows of R Aqr was done by \citet{Hollis1999} 
using a Fabry-Perot imaging spectrometer, providing a full field of view spectral coverage in the \hbox{[N\,{\sc ii}] 6584 \AA} emission line. In agreement with \citet{Solf1985}, their data and modelling confirms the outer bipolar nebula seen at an inclination angle $i\sim72$\textdegree\ and whose waist is expanding with a velocity of 55 km s$^{-1}$. However, contrary to the first modelling, instead of the inner nebula, they find a better fit to the data with a two-sided collimated jet emanating from the central star in the northeast (NE) and southwest (SW) direction, resembling a small-scale helical structure forming around a larger scale helical path probably due to the precession (see also their figure 1 presenting the model as an online animation). Radial velocities of this jet are found to be $\pm$175 km s$^{-1}$. They consider the long-slit data of \citet{Solf1985} limited, compared to their Fabry-Perot data cube, and therefore consider the inner nebula interpretation a possible mistake and/or confusion by the evolving jet at the time of the particular observations.

The subsequent published observational data at small scales (e.g. \citealt{2004A&A...424..157M}, \citealt{2007ApJ...660..651N}, \citealt{2017A&A...602A..53S}, \citealt{Bujarrabal2021}) and large scales  (e.g. \citealt{Kellogg2001}, \citealt{Liimets2018}, \citealt{Melnikov2018}) are in line with the model of an expanding outer hourglass nebula and an S-shaped jet with a complex morphology launched from the central binary. 
In the early years, it was observed that the northern part of the jet is mostly blue-shifted and the southern part red-shifted  (\citealt{Solf1985}, \citealt{Hollis1999}). However, high resolution integral field unit observations of the \hbox{[O\,{\sc iii}] 5007 \AA} emission line taken during 2012 \citep{Liimets2018}, where the jet is the most prominent, show that the northern part is predominantly red-shifted and the southern part blue-shifted, with radial velocities ranging from about $-60$ to $+140$ km s$^{-1}$. This discrepancy is attributed to generally very complex line profiles present in various features of the jet, a fact that is pointed out also by the above mentioned works, as well as the higher spectral and spatial resolution of their data. \citet{Liimets2018} report large morphological and brightness changes in the evolution of individual jet features during the time span of more than two decades. Furthermore, fast moving features, perpendicular to the overall expansion direction are found. 
They conclude that the interpretation of the R Aqr jet is complex, requiring considering ionisation variations and illumination effects in addition to the matter simply moving away from the central star.

In this paper we present a 3D model view of the nebular features of R~Aqr based on the analysis of high-dispersion spectra. Our data are interpreted by means of a proposed morpho-kinematic model to unveil the formation mechanism of the nebula associated with this symbiotic system. This paper is organised as follows. In Section~\ref{sec:obs} we describe our observations and their processing. Section~\ref{sec:model} introduces our morpho-kinematic model. Its consequences are discussed in Section~\ref{sec:discussion} and our conclusions and final remarks are presented in Section~\ref{sec:summary}.



\section{Observations and data preparation}
\label{sec:obs}


Long-slit, high resolution spectroscopic observations of R\,Aqr were obtained using the Manchester Echelle Spectrometer \citep[MES;][]{Meaburn} mounted on the 2.1m telescope at the Observatorio Astron\'{o}mico Nacional in San
Pedro M\'{a}rtir (SPM, Ensenada, Mexico)\footnote{The Observatorio Astronómico 
Nacional at the Sierra de San Pedro Mártir (OAN-SPM) is operated by the 
Instituto de Astronomía of the Universidad Nacional Autónoma de México.}. The SPM MES observations of R\,Aqr were conducted during two epochs, on 2022 Oct 24-26 and 2023 July 29-30. 

\begin{table}
\caption{Details of the SPM-MES observations of R\,Aqr. The offsets are denoted with respect to the position of the central star.}
\label{slits}
\centering
\setlength{\tabcolsep}{0.8\tabcolsep}   
\label{tab:obs}
\begin{tabular}{crccccc} 
\hline
Slits &   Offset  &  Filter   & PA         & Date         & Exp. Time & Seeing\\
      &   (arcsec)  &           &($^{\circ}$) &             & (s)  & (arcsec)    \\
\hline
S1 & 0.0 & H$\alpha+$[N\,{\sc ii}] & 0 &  2022 Oct 24 & 300 & 2.0 \\
S2 & 12.0 & H$\alpha+$[N\,{\sc ii}] & 0 &  2022 Oct 24 & 300 & 2.0 \\
S3 & 25.0 & H$\alpha+$[N\,{\sc ii}] & 0 &  2022 Oct 24 & 300 & 2.0 \\
S4 & 37.0 & H$\alpha+$[N\,{\sc ii}] & 0 &  2022 Oct 24 & 300 & 2.0 \\
S5 & 52.0 & H$\alpha+$[N\,{\sc ii}] & 0 &  2022 Oct 24 & 300 & 2.0 \\
S6 & 69.0 & H$\alpha+$[N\,{\sc ii}] & 0 &  2022 Oct 24 & 300 & 2.0 \\
S7 & $-$14.0 & H$\alpha+$[N\,{\sc ii}] & 0 &  2022 Oct 24 & 300 & 2.0 \\
S8 & $-$29.0 & H$\alpha+$[N\,{\sc ii}] & 0 &  2022 Oct 24 & 300 & 2.0 \\
S9 & $-$46.0 & H$\alpha+$[N\,{\sc ii}] & 0 &  2022 Oct 24 & 300 & 2.0 \\
S10 & $-$59.0 & H$\alpha+$[N\,{\sc ii}] & 0 &  2022 Oct 24 & 300 & 1.9 \\
S11 & 0.0 & H$\alpha+$[N\,{\sc ii}] & 90 &  2022 Oct 26 & 300 & 2.0 \\
\hline
S1$'$ & 6.0 & [O\,{\sc iii}] & 0 &  2023 Jul 29 & 1200 & 1.7 \\
S2$'$ & 0.0 & [O\,{\sc iii}] & 0 &  2023 Jul 29 & 1200 & 1.7 \\
S3$'$ & $-$6.0 & [O\,{\sc iii}] & 0 &  2023 Jul 29 & 1200 & 1.7 \\
S4$'$ & 0.0 & [O\,{\sc iii}] & 20 &  2023 Jul 29 & 1200 & 1.7 \\
S5$'$ & 0.0 & [O\,{\sc iii}] & 30 &  2023 Jul 29 & 1200 & 1.7 \\
S6$'$ & 0.0 & [O\,{\sc iii}] & 40 &  2023 Jul 30 & 1200 & 1.7 \\
S7$'$ & 0.0 & [O\,{\sc iii}] & 50 &  2023 Jul 30 & 1200 & 1.8 \\
\hline
\end{tabular}
\end{table}

\begin{figure*}
\begin{center}
\includegraphics[angle=0,width=1.0\linewidth]{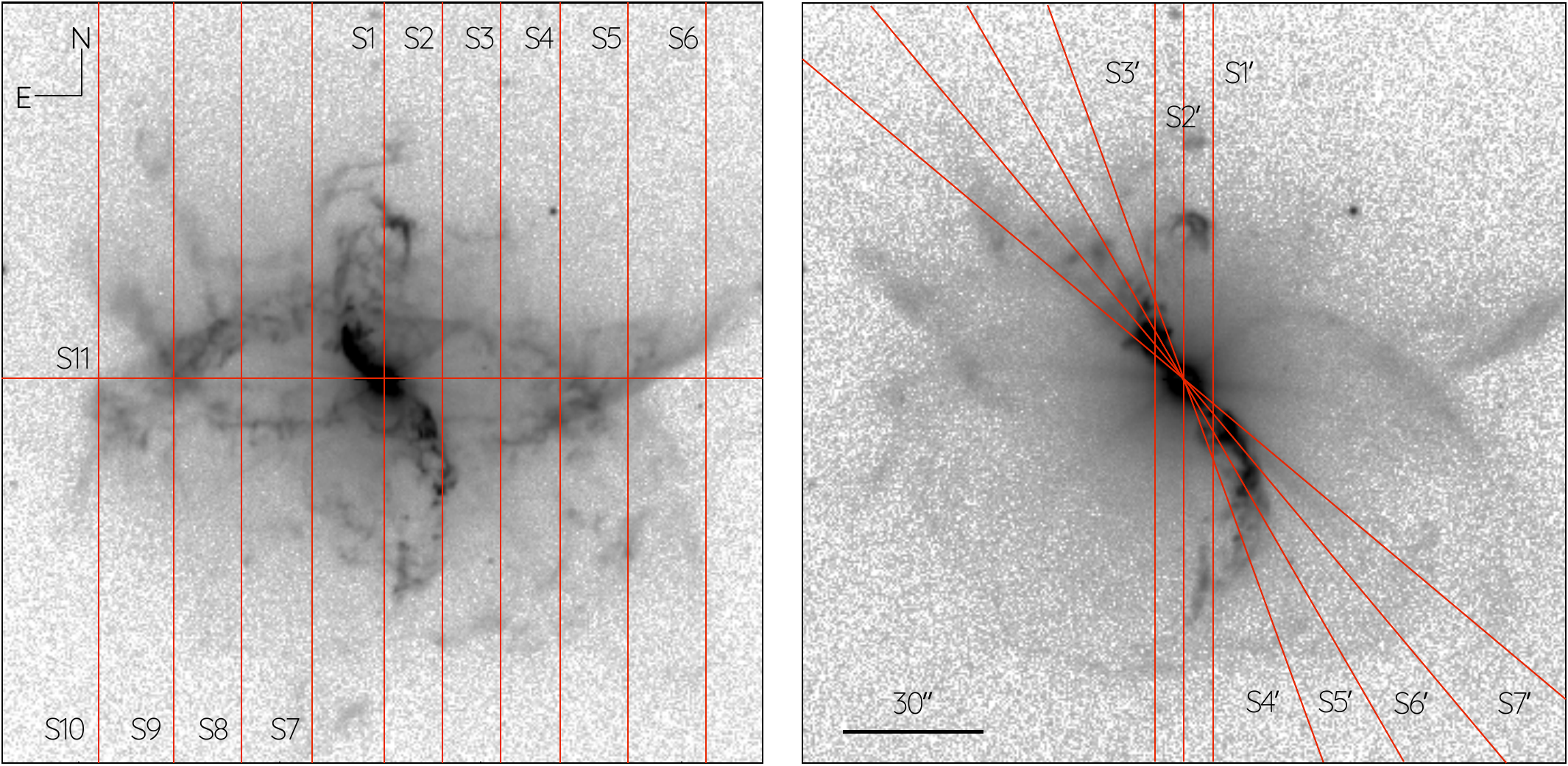}
\caption{
Positions of the SPM MES slits (red) overlaid on optical images of R\,Aqr. 
The left and right panels show the position of the H$\alpha+$[N~{\sc ii}] and [O\,{\sc iii}] slits overlaid on logarithmic scaled images obtained on the same emission lines by \citet{Liimets2018}.}
\label{fig:slits}
\end{center}
\end{figure*}

We used the E2V 42-40 CCD with pixel size 13.5 $\mu$m with a 4$\times$4 on-chip binning, resulting in a plate scale of 0.702 arcsec pix$^{-1}$. 
A total of 18 positions were used to acquired high-resolution SPM MES spectra of different morphological features in the nebula associated with R\,Aqr. 
Observations were obtained in two spectral ranges.  
The first run used a filter centred on the H$\alpha$ emission line ($\Delta \lambda=90$ \AA) to isolate the 87th echelle order, which also includes the [N\,{\sc ii}]~6548,6584~\AA\ doublet. 
The spectral scale is 0.10 \AA~pix$^{-1}$. 
Motivated by the results presented in \citet{Liimets2018} where the jet-like feature is best mapped by the [O\,{\sc iii}]~5007~\AA\, emission, we obtained a second set of spectra with the interference filter $\lambda_{\rm c}=5020$ \AA  ~($\Delta \lambda=70$ \AA) that includes the emission line of [O\,{\sc iii}]$\lambda5007$ \AA ~to isolate the 114th echelle order (0.086 \AA~pix$^{-1}$ spectral scale). A slit width of 150 $\mu$m  (1.9 arcsec) was used, resulting in a spectral resolution of $\simeq 12\pm1$ km s$^{-1}$. Further details of the observations are listed in Table~\ref{tab:obs} and the slit positions are illustrated on Fig.~\ref{fig:slits}. Spectra extracted using the H$\alpha$+[N\,{\sc ii}] filter are labelled S1--S11 and those obtained using the [O\,{\sc iii}] filter correspond to S1$'$--S7$'$.

\begin{figure*}
\begin{center}
\includegraphics[angle=0,width=.9\linewidth]{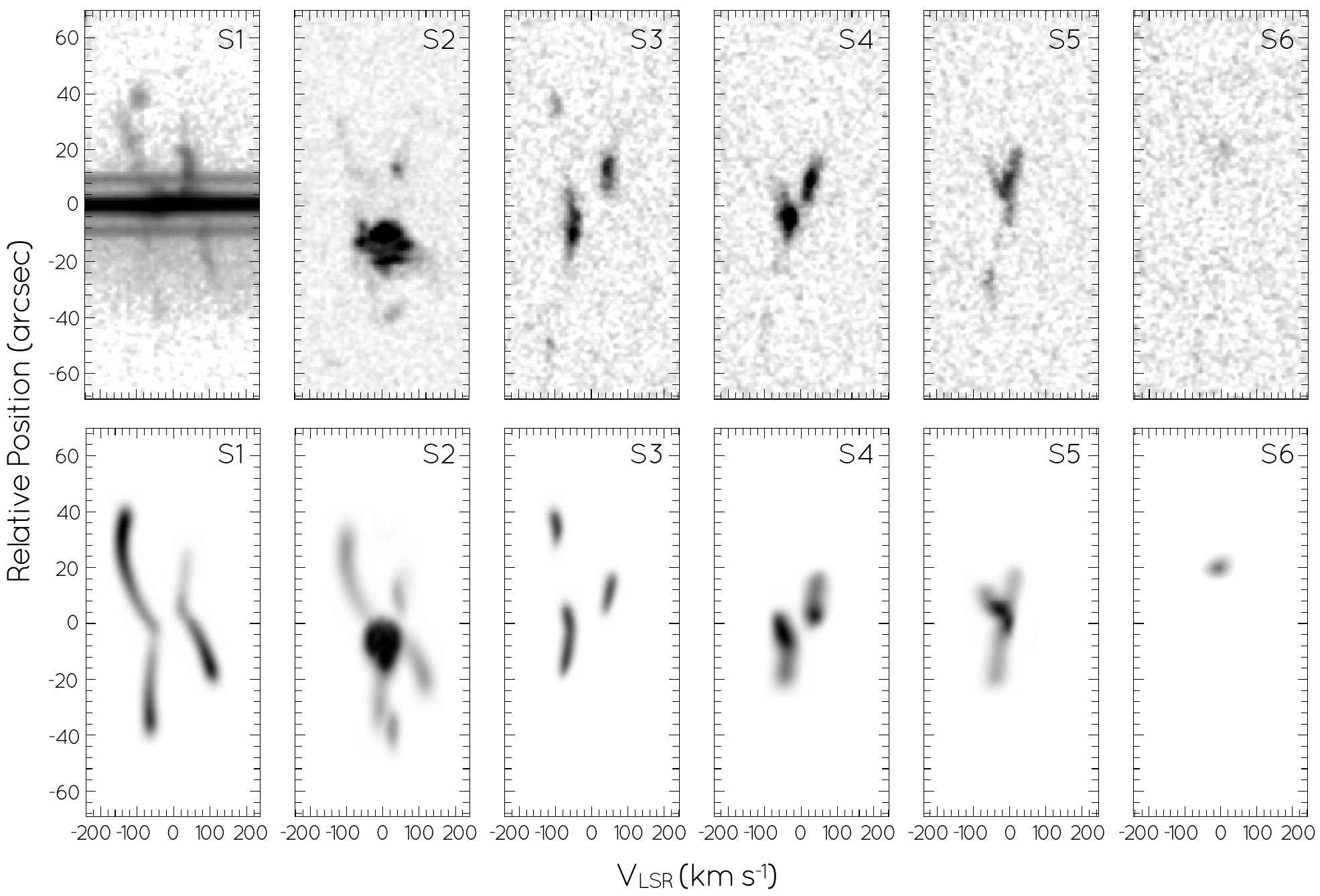}
\caption{
Top panels: 
PV diagrams obtained from the [N\,{\sc ii}] 6584 \AA\ emission line from slits S1 to  S6. 
Bottom panels: Synthetic PV diagrams (in logarithmic scale) obtained from our best {\sc shape} morpho-kinematic model.}
\label{fig:pv_ha}
\end{center}
\end{figure*}

\begin{figure*}
\begin{center}
\includegraphics[angle=0,width=0.75\linewidth]{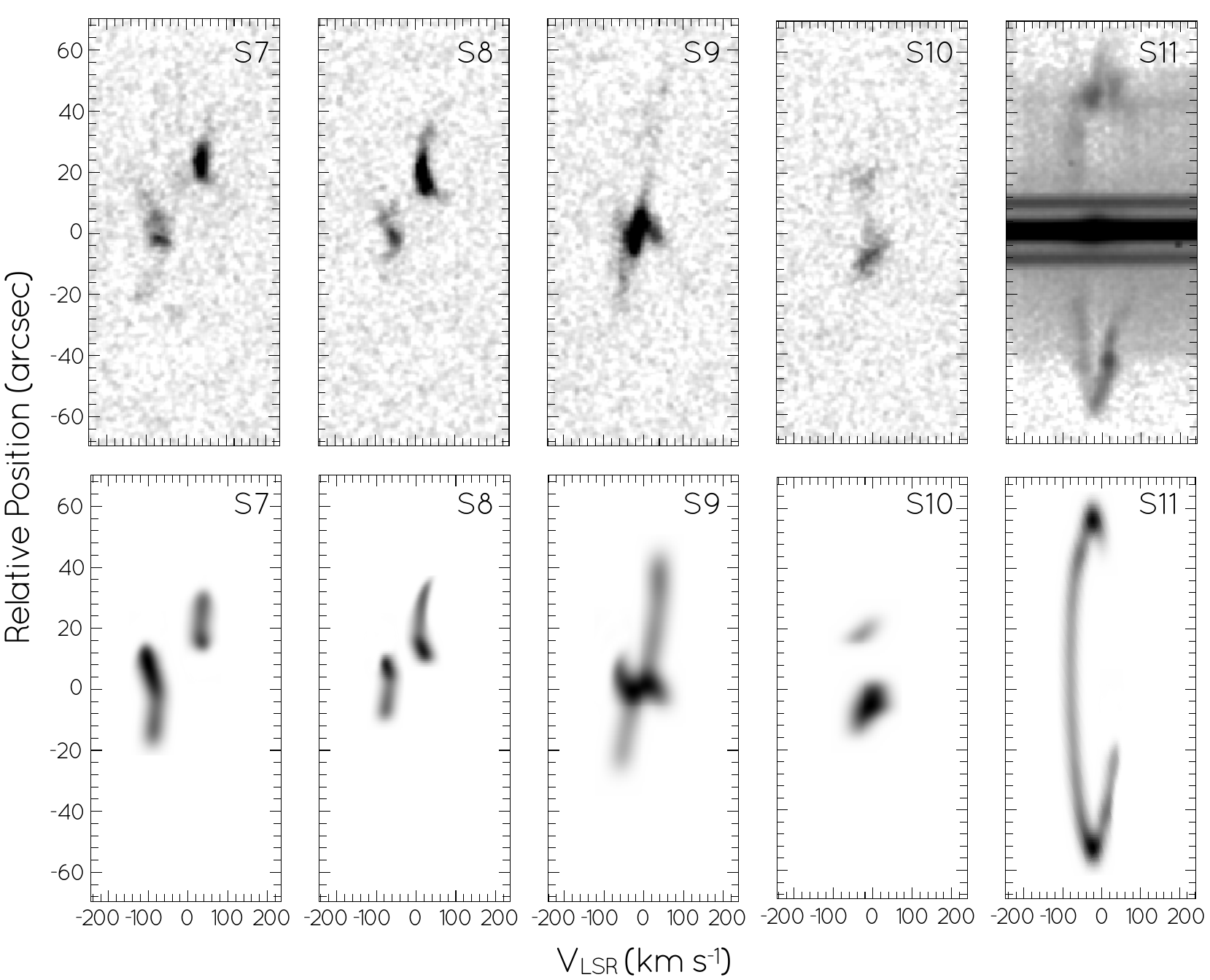}
\caption{Same as Fig.~\ref{fig:pv_ha} but for slits S7 to S11. 
}
\label{fig:pv_ha_2}
\end{center}
\end{figure*}

The spectra were processed using standard calibration routines in {\sc iraf} \citep{Tody}. This included bias subtraction and wavelength calibration with ThAr arc lamps obtained immediately before and after the science observations. The wavelength accuracy is estimated to be $\pm$1 km s$^{-1}$.

\begin{figure*}
\begin{center}
\includegraphics[angle=0,width=1\linewidth]{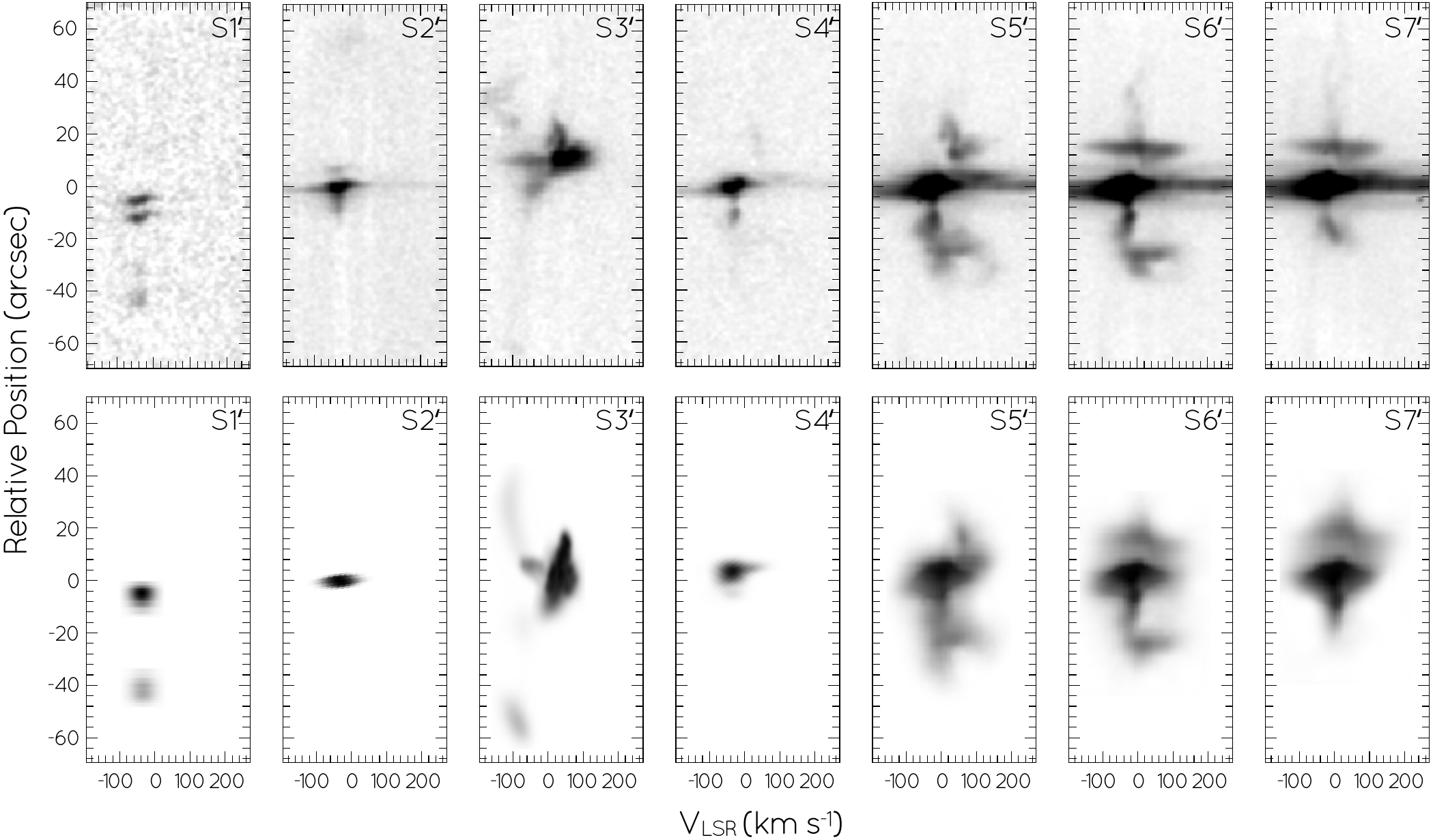}
\caption{Same as Fig.~\ref{fig:pv_ha} but for the [O\,{\sc iii}] observations obtained from slits S1$'$--S7$'$. The bottom panels show their corresponding synthetic counterparts obtained from our best morpho-kinematic model obtained with {\sc shape}.}
\label{fig:pv_oiii}
\end{center}
\end{figure*}

Examples of position-velocity (PV) diagrams from slits S1--S6 and S7--S11 are presented in the top rows of Fig.~\ref{fig:pv_ha} and \ref{fig:pv_ha_2}, respectively. These correspond to PV diagrams of the [N\,{\sc ii}]~6584~\AA. We note that PVs obtained from the H$\alpha$ emission (not shown here) are very similar to those of the [N\,{\sc ii}], but the former are affected by thermal broadening. Those obtained from slits S1$'$--S7$'$ showing the [O\,{\sc iii}]~5007~\AA\, emission are presented in Fig.~\ref{fig:pv_oiii}. Radial velocities were corrected for the local standard of rest (LSR), and then we derived a systemic velocity of $V_\mathrm{sys}=-$22.6 km s$^{-1}$.

\section{Morpho-kinematic model}
\label{sec:model}

The morpho-kinematic model was produced using the {\sc shape} software \citep{Steffen2011} version 5.0. 
This software allows the user to define relatively simple geometrical morphologies that can be characterised by expansion velocity patterns. 
Ultimately, {\sc shape} allows to produce synthetic images and synthetic spectra as well by defining synthetic slits that can be directly compared with images and PV diagrams.

The process to deterine the 3D structure of a source is iterative and can be summarised in the following steps: 
\begin{enumerate}
\item An initial structure (or primitive object) with certain properties (size, 3D velocity expansion pattern, density) is proposed through ``modifiers''; 
\item one or a number of synthetic slits (similar as those used in the observations) are defined to extract synthetic spectra and produce synthetic PVs; 
\item the synthetic images and PVs are compared with the observed ones; 
\item the structure is iteratively edited to produce a better fit of the observations. 
\end{enumerate}
\noindent A good model is that capable of reproducing the largest number of observations.  
Through this process, the distance is adopted to be $d$=180~pc (see Section~\ref{sec:introduction}).

{\sc shape} allows the user to define different expansion velocity laws for the morphological structures. This work adopts expansion velocity functions of the form
\begin{equation}
    \varv(r) = V_{0}~\left(\frac{r}{R_0}\right),
\end{equation}
\noindent where $V_0$ and $R_0$ are constants. Here $r$ denotes the real 3D distance of each point on a structure measured from the origin, that is, $\varv(r)$ is the true expansion velocity from the centre\footnote{We note that adopting this velocity function assumes that all nebular features present a homologous expansion.}. The errors have been estimated by varying the morphological features and comparing the synthetic spectra with those obtained from the observations.

Our best model requires three main elements, which we have named (a) large (outer) hourglass, (b) bipolar structure and (c) spiral-like filament. All synthetic PV diagrams extracted from the model are compared to their corresponding positions obtained from the observations in the bottom rows of Figs.~\ref{fig:pv_ha}, \ref{fig:pv_ha_2} and \ref{fig:pv_oiii}.

It is important to note that in order to produce the observed brightness distribution, ad hoc density variations need to be considered for each structure in the {\sc shape} model. The density in {\sc shape} is not meant to reproduce the physical density parameter, but the spatially-varying emission intensity of each structure, i.e., it has no physical meaning. To illustrate the brightness variations between different morphological features in the nebula around R~Aqr, we show in Appendix~\ref{sec:app} the nebular H$\alpha$+[N\,{\sc ii}] image obtained with the VLT FORS2 instrument in arbitrary flux units.

Fig.~\ref{fig:model} shows all of the structures required by the best {\sc shape} model. 
The top-left panel presents the required orientation fitting the observations, while the top-middle and top-right panels show edge-on and pole-on views. 
Finally, the bottom-left panel of this figure presents a synthetic image of the model that is compared with the optical H$\alpha$+[N\,{\sc ii}] image obtained at the VLT shown in the bottom-right panel. Further details of the three morphological components are given in the following subsections.

\subsection{Large (outer) hourglass}

\begin{figure*}
\begin{center}
\includegraphics[angle=0,width=1.0\linewidth]{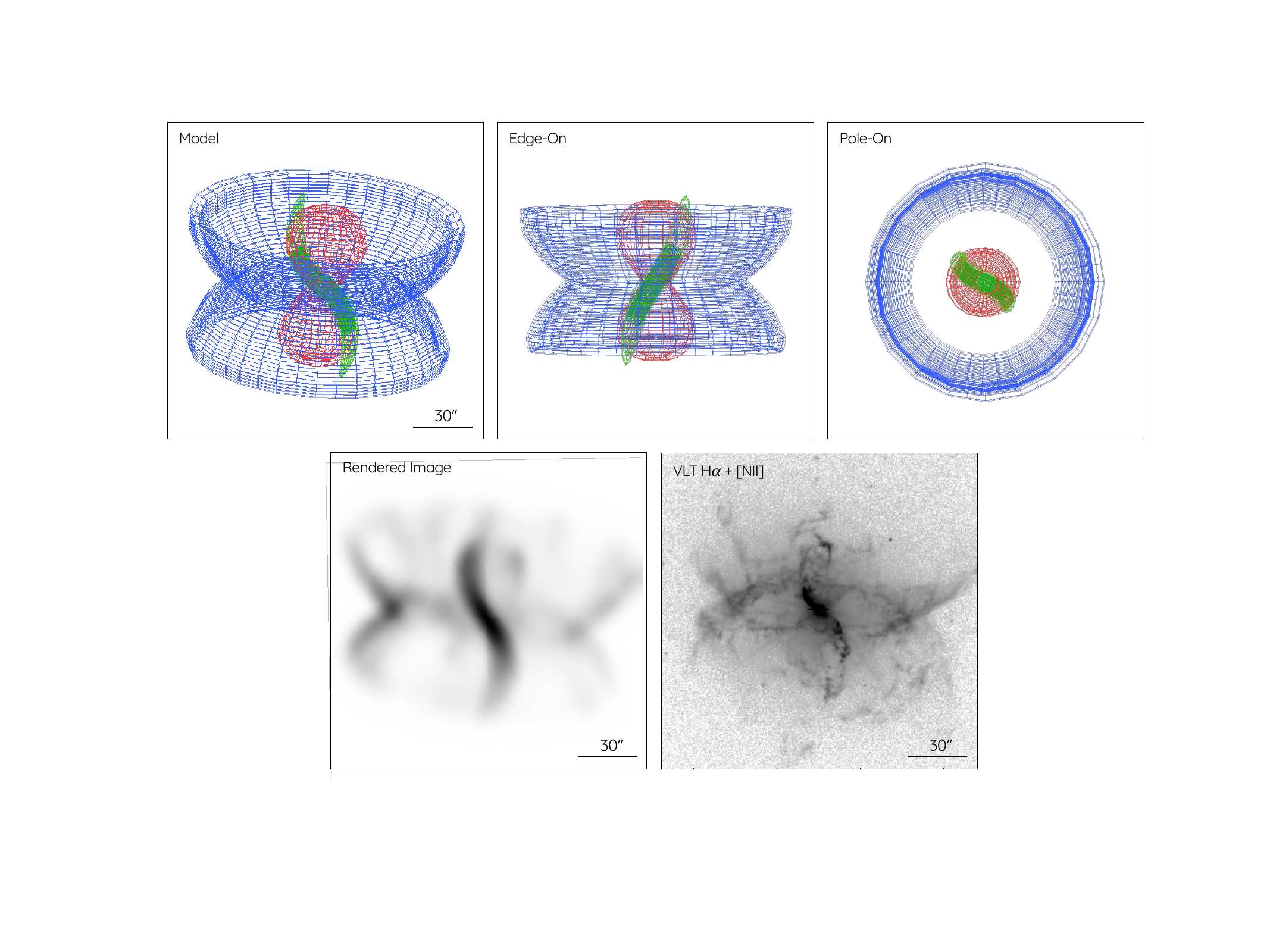}
\caption{{\bf Top:} Morpho-kinematic model of R\,Aqr obtained with the software {\sc shape}. Different coloured mesh represent the three main components of our model: the large hourglass in blue, the inner bipolar in red and the spiral-like filament in green (see Section~\ref{sec:model} for details). The panels show our best model through different viewing angles. {\bf Bottom:} left-hand panel shows a rendered image obtained from our {\sc shape} model which is compared with the VLT H$\alpha+$[N\,{\sc ii}] image (bottom right panel).}
\label{fig:model}
\end{center}
\end{figure*}

The {\sc shape} modelling started by proposing a 
bipolar element to model the external hourglass nebula around R~Aqr. 
This structure has been squeezed at its equatorial plane 
to produce a narrow waist. 
A model with a waist of 45.6~arcsec in radius reproduces best the spectra and images. 
The radius at the North 
opening of this structure is adjusted to 72.5~arcsec, but the South 
counterpart has an opening of 75.7~arcsec in radius. 
The total extension along the N-S direction of this bipolar structure 
is 79.8~arcsec (39.9 arcsec for each lobe). 
Its symmetry axis is tilted 18$\pm3^\circ$ with respect to the plane of the sky\footnote{That is, $72\pm3.0^\circ$ between its symmetry axis and the line of sight.} along a PA of $-5^\circ$ with its Northern end pointing toward us. This large (outer) bipolar structure 
is shown as a blue mesh in Fig.~\ref{fig:model}.

The observations are best reproduced by using an expansion velocity law of
\begin{equation}
    \varv_1(r) = 170 \left(\frac{r}{90''} \right) \mathrm{km~s}^{-1},
\end{equation}
\noindent that implies a kinematic age $\tau_1$=450$\pm$25~yr after accounting for uncertainties.  
At the waist ($r = 45.6$ arcsec), the real expansion velocity is 86$\pm$5~km~s$^{-1}$, whereas any point on the bottom circumference defined by the opening of the southern lobe ($r\approx78.3$ arcsec) has a real expansion velocity of $\approx$148 km~s$^{-1}$. 
%



\subsection{Bipolar structure}

The inner bipolar structure (red mesh in Fig.~\ref{fig:model}) has the same inclination properties as the outer hourglass described in the previous section, and has thus been modelled similarly, but with a structure including two closed lobes. 
The bipolar lobes of the best model have an equatorial waist with size 7.7~arcsec (3.9~arcsec in radius), a maximum width of 31.8~arcsec (15.9~arcsec in radius), and a total height of 39.9~arcsec each.

This best model resulted in an expansion velocity law 
\begin{equation}
    \varv_2(r) = 150\left(\frac{r}{47''}\right)~\mathrm{km~s}^{-1},
\end{equation}
\noindent that implies a kinematic age $\tau_2$=270$\pm$20~yr after accounting for uncertainties.  
The expansion velocity at the bipolar waist is 12.3$\pm$2~km~s$^{-1}$, whereas at the furthermost point of each lobe, i.e. their polar position, the true 3D velocity is 125$\pm$10~km~s$^{-1}$.

\subsection{Spiral-like filament}

A third structural component is needed in order to fit the bright central spiral-like filament of R\,Aqr. 
For practical reasons, this spiral-like filament was modelled using three twisted cylindrical structures that are divided into a central one and two at the farthermost locations (with respect to the origin). These three structures are illustrated as green meshes in Fig.~\ref{fig:model}.

From a pole-on view (top right panel of Fig.~\ref{fig:model}), the whole spiral-like filament has a pitch angle of 35$^\circ$. The model also required this structure to have a total elevation of 41~arcsec from the origin. We note that this is only slightly larger than the height of a lobe of the inner bipolar structure. That means that the tips of the spiral-like filament are located at a distance of $r$=43.5 arcsec from the origin. 

The three components of the spiral-like filament were modelled with a velocity law
\begin{equation}
    \varv_\mathrm{3}(r) =  120 \left(\frac{r}{40''}\right)~\mathrm{km~s}^{-1}.
\end{equation}
\noindent implying a kinematic age of $\tau_3$=285$\pm$20~yr.

\section{Discussion}
\label{sec:discussion}

\begin{figure}
\begin{center}
\includegraphics[angle=0,width=1.0\columnwidth]{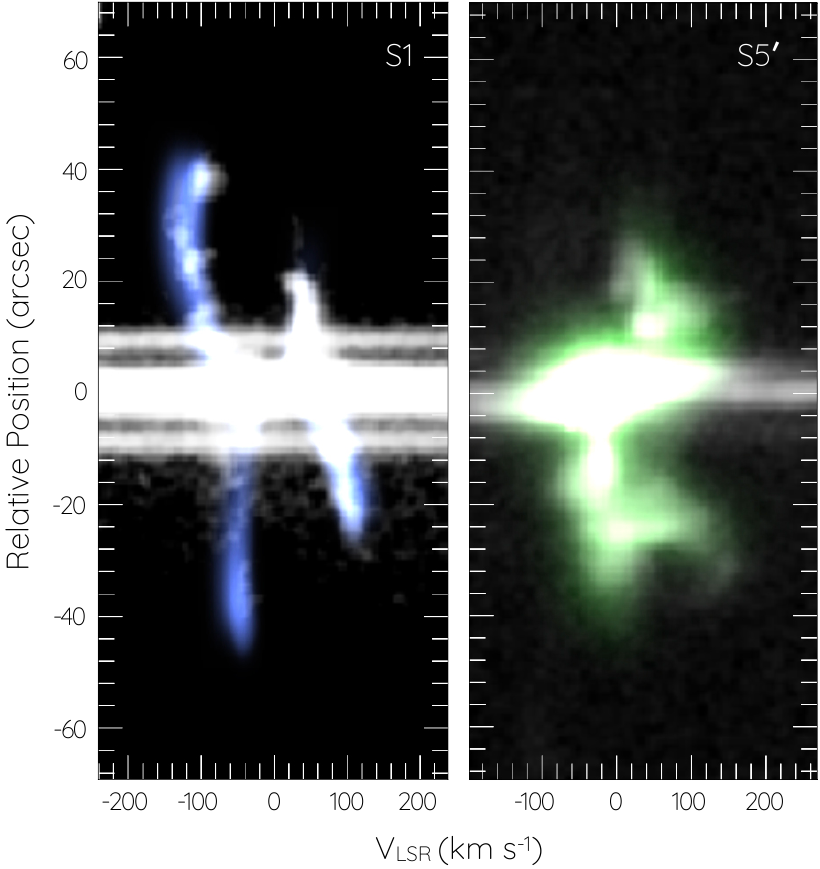}
\caption{Contribution of the different kinematic components in our {\sc shape} model for slits S1 and S5$'$. The panels show in blue and green the contributions from the outer hourglass and spiral-like filament structures in the R\,Aqr {\sc shape} model.}
\label{fig:s}
\end{center}
\end{figure}

This is the first time that three different structures are needed in order to explain the properties of the nebula associated with R Aqr. \citet{Solf1985} suggested the presence of the inner and outer hourglass structures, while \citet{Hollis1999} required a precessing jet in addition to the outer hourglass. 
We note that \citet{Solf1985} suggested that the PV diagram extracted with PA=0$^{\circ}$ at the central region of R~Aqr to be produced by the inner bipolar structure, however, this PV is largely reproduced by the characteristics of the outer (large) hourglass structure. This is illustrated in the left panel of Fig.~\ref{fig:s}. Most of the bright inner structure is saturated which hampers a good interpretation using slit S1. Nevertheless, a comparison between the contribution of the spiral-like filament and the PV extracted from slit S5$'$ is presented in the right panel of this figure. 

Given that it is located at a relatively short distance, R~Aqr is a one of a kind among symbiotic stars. This allows us to resolve in detail all the morphological features produced by the complex interactions of its binary system at its core. These large scale structures have to be a consequence of what we expect is occurring at the binary's orbit scales, and accretion physics is to be blamed for all of it.

It is accepted that a WD is accreting material from a red giant star in symbiotic systems. 
The material cannot fall directly into the compact component and thus, an accretion disk should be formed. 
In fact, recently \citet{Merc2024} suggested that at least 80 per cent of symbiotic systems should have an accretion disk. This fact seems to be true regardless of the physics behind the accretion, which could be produced by different mechanisms \citep[Bondi-Hoyle, Roche Lobe overflow or a wind Roche lobe overflow;][]{Bondi1944,Podsiadlowski2007}.
Numerical simulations predict that as a consequence of these interactions, the density distribution surrounding the symbiotic system should have a toroidal-like morphology \citep[e.g.,][]{Makita2000,deVal2009,Liu2017,Lee2022}.

According to the jet-feedback mechanism, jets might be created as a result of accretion onto a compact object \citep{Soker2016}.
Jets have been routinely detected from some symbiotic systems \citep[see for example][for the case of HM Sge]{Corradi1999,Toala2023}, but not from all of them. \citet{Soker2000} suggested that in order to launch a jet, the compact object should accrete material at a rate above a certain limit.
One would expect the jet to be launched towards a direction (more or less) perpendicular to the orbital plane and toroidal structure.  
That is, whenever present, jets should be able to carve bipolar nebulae around symbiotic systems. A situation that has been proven to be occurring in R Aqr.

Evidence of the presence of a precessing jet at the core of R Aqr has been presented in several works using different wavelengths, and it is indeed more or less oriented in the N-S direction with a certain inclination \citep[see][and references therein]{Melnikov2018,Sacchi2024}.
In our best morpho-kinematic model, the effects of the jet are unveiled by the spiral-like filament. We suggest that this structure is tracing the location of the interaction of the precessing jet material impinging into the circumstellar material. The fact that the spiral-like filament is at the edge of the inner bipolar seems to suggest that the latter is being carved by the jet action. A suggestion that is supported by both structures, the inner bipolar and the tips of the spiral-like filament, having virtually the same kinematic ages of about 280~yr. 
The high velocity of the jet produces X-ray-emitting, adiabatically-shocked gas with high pressure that shows regions of interactions of the jet with the circumstellar material as revealed by {\it Chandra} X-ray observations \citep[][]{Kellogg2001,Kellogg2007,Toala2022}\footnote{See also the Chandra Image Gallery entry of R Aqr (\url{https://chandra.harvard.edu/photo/2017/raqr/}).}.


We note that it is very likely that the small bipolar might not be as simple and smooth as our {\sc shape} model suggest. A detailed inspection of the exquisite quality VLT images and publicly available {\it HST} images of R Aqr suggest that in fact this structure might be dominated by the presence of blister-like structures. However, the proposed model seems to be good enough to reproduce the SPM MES data, after accounting for their spatial resolution.

\begin{figure*}
\begin{center}
\includegraphics[angle=0,width=\linewidth]{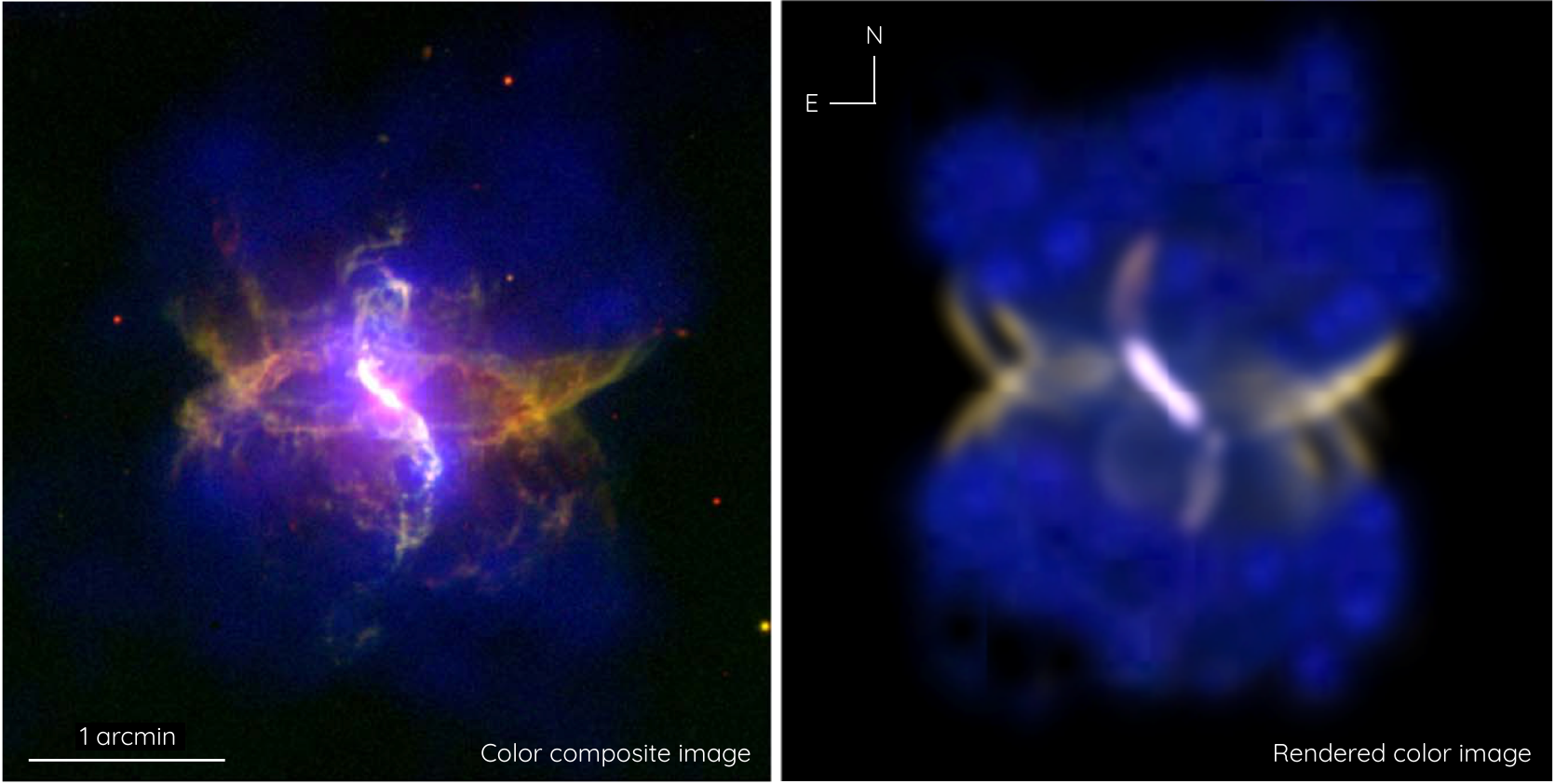}
\caption{(left) Colour-composite images of R\,Aqr from \citet{Toala2022}. Green and red correspond to the 
[O~{\sc ii}] and H$\alpha+$[N\,{\sc ii}] narrowband images obtained with VLT and presented in \citet{Liimets2018}, the soft (0.3–0.7 keV) X-ray emission detected by XMM-Newton EPIC is presented in blue. (right) {\sc shape} rendered colour image.}
\label{fig:xray}
\end{center}
\end{figure*}

The best {\sc shape} model suggests that the kinematical age of the outer hourglass nebula is 450$\pm$25~yr, which is about 30 per cent smaller than previous estimates that range between 600 and 650~yr \citep[see][and references therein]{Solf1985,Liimets2018}. 
The distance is not to be blamed for the differences, but the accurate determination of the velocity of the different morphological features. The SPM MES observations used here are the highest spectral resolution observations of the nebula around R Aqr discussed in the literature thus far and their interpretation through the {\sc shape} model resulted in the most accurate 3D velocity mapping. As mentioned above, the model presented here adopts homologous expansion laws ($\varv \propto r$) for the different morphological features. The assumption behind this is that there has been no change in the velocity of each particle since its ejection.

The determination of the kinematic signatures of the different structures in R Aqr can help us peer into the formation mechanism of its associated nebula. If one accepts that the current precessing jet has indeed produced the inner bipolar structure, a similar idea might apply to the outer hourglass nebula. If the outer hourglass nebula was also carved by the actions of a primordial jet, its ejection ceased for about 170~yr, when the current precessing jet started acting. We note that similar ideas were proposed by \citet{Toala2022}. That work proposed that the extended X-ray emission detected by {\it XMM-Newton} is the reminiscent of a previous bipolar ejection. After this primordial jet stopped its action, the outer nebula expanded given the high pressure of its contained hot gas.

This idea has some consequences for other X-ray-emitting symbiotic systems. For example, accretion dominated systems classified as $\delta$-type X-ray-emitting symbiotic stars are known to harbour highly extinguished, high temperature plasma. This is accepted to be located at the boundary region between the WD component and the accretion disk and it is usually detected in hard X-rays with energies $E>$3.0 keV \citep{Luna2013}. However, in some cases $\delta$-type X-ray-emitting symbiotic systems suddenly develop soft X-ray emission in the 0.3-3.0 keV energy range becoming $\beta/\delta$-type sources \cite[e.g., the case of T~CrB; see][and references therein]{Zhekov2019,Toala2024_TCrB}. Some works have attributed this behaviour to a disk instability and/or a change in the absorbing column density. But recent work by \citet{Toala2024} instead suggests that the appearance of the soft X-ray emission is due to the creation of a hot bubble, very likely produced by a jet as clearly seen in R~Aqr and other sources such as CH Cyg and HM Sge \citep[see][]{Toala2023,Toala2023CHCyg}. We further used the morpho-kinematic model obtained with {\sc shape} to illustrate the presence of the hot, X-ray-emitting material in R~Aqr in Fig.~\ref{fig:xray}.






\section{Summary}
\label{sec:summary}

We present here the analysis of high-dispersion SPM MES observations of the nebula associated with the symbiotic system R~Aqr. H$\alpha$+[N\,{\sc ii}] and [O\,{\sc iii}] spectra observations were interpreted by means of the {\sc shape} software to produce a morpho-kinematic model of the nebula. Synthetic images and PV diagrams were compared to those obtained from observations to assess the best model.

It was found that the best morpho-kinematic model of the nebula associated with R~Aqr requires three main components. An outer hourglass, a bipolar structure located inside the large structure and a spiral-like filament that is entwined on the edge of the bipolar. We found that the corresponding kinematic ages of these structures are $\tau_1$=450$\pm$25 for the outer hourglass, $\tau_2$=270$\pm$20 for the inner bipolar and $\tau_3$=285$\pm$20~yr for the spiral-like filament.

We suggest that the spiral-like filament is the signature of the interaction of the current precessing jet with the circumstellar material ejected by the symbiotic system at the heart of R~Aqr. We propose that the action of the jet is simultaneously producing the inner bipolar structure. The later is confirmed by their very similar kinematical ages ($\approx$280~yr). Following this idea, we suggest that a previous jet carved the most extended hourglass structure. Which means that R~Aqr stopped producing a jet for about 170~yr. The intermittent activation of the jet might help explaining the sudden production of soft X-ray emission in some other X-ray-emitting symbiotic systems.

\section*{Acknowledgements} 

The authors thank the referee V. Bujarrabal for comments and suggestions that helped improve the presentation and discussion of the results. ES thanks UNAM DGAPA for a postdoctoral fellowship. JAT thanks support from UNAM DGAPA PAIIT project IN102324. The Astronomical Institute of the Czech Academy of Sciences is supported by the project RVO:67985815. 
M.A.G.\ acknowledges financial support from grants CEX2021-001131-S funded by MCIN/AEI/10.13039/501100011033 and PID2022-142925NB-I00  from the Spanish Ministerio de Ciencia, Innovaci\'on y Universidades (MCIU) cofunded with FEDER funds.
MB y LS acknowledge support from UNAM DGAPA PAIIT project IN110122. GRL acknowledges support from Consejo Nacional de Ciencia y Tecnolog\'{\i}a (CONACyT) grant 263373 and Programa para el Desarrollo Profesional Docente (PRODEP, Mexico). This work has made an extensive use of NASA's Astrophysics Data System (ADS).

\section*{Data availability}

The data and morpho-kinematic model presented in this work are available in the article. They will be shared on reasonable request to the corresponding author.








\appendix

\section{Nebular Image of R~Aqr}
\label{sec:app}

\begin{figure*}
\begin{center}
\includegraphics[angle=0,width=\linewidth]{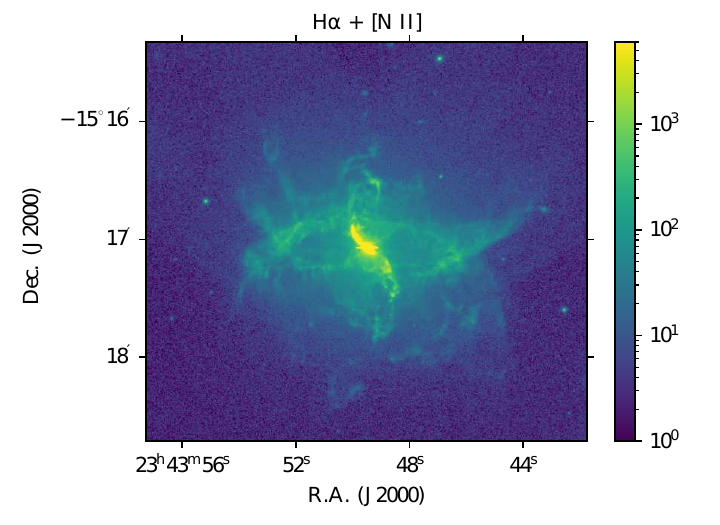}
\caption{Nebular image of R~Aqr obtained with the H$\alpha$+[N\,{\sc ii}] filter of the VLTO FORS2 instrument originally presented in \citet{Liimets2018}. The scale bar shows an arbitrary flux intensity to illustrate intensity values within the different morphological features in the nebula.}
\label{fig:flux}
\end{center}
\end{figure*}

Fig.~\ref{fig:flux} shows the nebular image of R Aqr obtained through the H$\alpha$+[N~{\sc ii}] filter of the VLT FORS2 instrument originally presented in \citet{Liimets2018}. This image is intended to show the reader the flux differences between the morphological features in the nebula. We note that the scale bar represents arbitrary units since the image is not flux-calibrated.



\bsp	
\label{lastpage}
\end{document}